\documentclass[pre,aps,twocolumn]{revtex4-1}
\usepackage[latin1]{inputenc}
\usepackage[english]{babel}
\usepackage{graphicx}
\usepackage{color}
\usepackage{amsmath}
\usepackage{ulem}
\usepackage{cancel}
\begin{document}

\title{\bf\Large Macroscopic amplification of electroweak effects in molecular Bose-Einstein condensates}
\author{P. Bargue\~no and F. Sols}
\affiliation{Departamento de F\'{\i}sica de Materiales, Universidad Complutense de Madrid, {\it E-28040}, Madrid, Spain}

\begin{abstract}
We investigate the possible use of Bose-Einstein condensates of diatomic molecules to measure
nuclear spin-dependent parity violation effects, outlining a detection method based
on the internal Josephson effect between molecular states of opposite parity. When applied to
molecular condensates, the fine
experimental control achieved in atomic bosonic Josephson junctions could provide data
on anapole moments and neutral weak couplings.
\end{abstract}
\maketitle

Since the discovery of the electroweak unification, parity violation (PV) has been one of the outstanding problems in
atomic and molecular physics. Although the success of the standard model (SM) of elementary particles is extraordinary,
the search for new physics is still carried out at high- and medium-energy particle colliders.
On the other hand,
very sensitive low-energy probes can also be implemented to test SM predictions \cite{Ginges2004}.
As compared to large accelerators, atomic experiments offer a complementary approach to study PV. In particular, nuclear spin-dependent PV effects focus on two causes. One
is the electroweak neutral coupling between electron vector- and nucleon axial-vector currents.
This can be parametrized by two constants, C$_{2u/d}$, describing the electron couplings to up/down quarks. These constants are the most poorly characterized parameters
in the SM \cite{Nakamura2010}. Thus, precise measurements of C$_{2u/d}$ are potentially sensitive to new physics
beyond the SM \cite{Langacker1992}. Another cause of spin-dependent PV is the
weak nucleon-nucleon interaction which gives place to the anapole moment \cite{Zeldovich1957,Flambaum1984}. This parity
violating magnetic moment, which results from the chirality acquired by the nucleon current \cite{comment}, couples to the spin
of the electron. As claimed by Haxton and Wieman \cite{Haxton2001}, the most practical strategy
for studying the effects of $Z^{0}$ exchange between hadrons is the investigation of the parity-violating
nucleon-nucleon interaction. One of its signatures is the resulting anapole moment, which has only been measured in
$^{133}$Cs \cite {Wood1997}. These high precision measurements allow for a better determination of
the meson-nucleon parity violating interaction constants, with standard reference values given by
the Desplanques, Donohue and Holstein (DDH) best values
\cite{Desplanques1980}. As there are
large uncertainties in these possible values (the DDH reasonable ranges), it is
therefore important to determine the weak coupling constants experimentally.

These and other important tests of the SM could be performed by using the ultra-high resolution
which is potentially available in cold molecule experiments. In particular, it is known that the internal structure and
specific properties of diatomic molecules can enhance the violation of some discrete symmetries, as
compared to their atom counterparts \cite{Flambaum1985}. In addition, heteronuclear molecules exhibit dipole-dipole interactions that
make them attractive candidates for use in quantum simulations of condensed-matter systems
\cite{Carr2009} and in quantum computation \cite{DeMille2002}.
Thus, experiments involving ultracold heteronuclear diatomic molecules
will be valuable to a large portion of the physics community. The fact that recently both homonuclear \cite{Danzl2010}
and heteronuclear \cite{Aikawa2010,{Ni2010}} ultracold molecules have been produced in the rovibrational ground state, suggests that the type of experiments envisaged here may not lie too far in the future.

The total nuclear spin-dependent parity violation effect is given by the Hamiltonian
\cite{Ginges2004}
\begin{equation}  \label{PV-Ham}
H_{\rm pv}=\frac{G_{F}}{\sqrt{2}}\kappa\frac{K}{I(I+1)}{\boldsymbol{\alpha}} \cdot {\bf{I}}\, \rho(r),
\end{equation}
where $G_{F}$ is the Fermi weak constant, ${\bf{I}}$ is the nuclear spin, $K=(I+1/2)(-1)^{I+1/2-l}$, $l$ is the
orbital angular momentum of the unpaired nucleon, ${\boldsymbol{\alpha}}$ is the vector whose components are
the Dirac matrices acting on the electron spinor, and $\rho(r)$ is the nuclear density.
The $\kappa$-term is
\begin{equation}
\kappa=\kappa_{a}-\frac{K-1/2}{K}\kappa_{2}+\frac{I(I+1)}{K}\kappa_{Q},
\end{equation}
where $\kappa_{a}$ stands for the anapolar term, $\kappa_{2}$ is due to the
electroweak neutral coupling between electron vector- and nucleon axial-vector currents
and $\kappa_{Q}$ represents
the interaction due the nuclear weak charge perturbed by the hyperfine interaction \cite{Ginges2004}.
We note that the nuclear anapole moment is the dominant source of PV in atoms with atomic number
$A\agt 20$ \cite{Haxton2001} since it increases as $\kappa_{a}\sim g\, A^{2/3}$, where $g$
is the strength of the weak interaction between the unpaired nucleon and the nuclear core
(the $\kappa_{Q}$ term also increases as $ A^{2/3}$ but the numerical coefficient is very small). Let us recall that there have been
proposals for measuring the anapole moment in the ground state of heavy alkali-metal atoms \cite{Gomez2007}.
Electroweak calculations show that, for a heavy nucleus such as, e.g., Rb, $\kappa_{a}/\kappa_{Q}=20$ and
$\kappa_{a}/\kappa_{2}=5$ \cite{Sheng2010}.

Concerning molecules, we note that the levels of opposite parity are four to five orders of magnitude closer in
diatomic molecules than in atoms, which results in
stronger mixing of those levels by PV \cite{Flambaum1985}.
The fact that the near-degeneracy of a pair of opposite parity levels enhances the level mixing
caused by the weak interaction can be exploited for precision measurements, as shown by the Berkeley group with atomic dysprosium \cite{Nguyen1997}.
We note that this enhancement mechanism does not work for homonuclear diatomic molecules
\cite{Khriplovichbook}.
Moreover, this energy interval between
opposite parity levels can be made even smaller by applying an external magnetic field. In addition, and contrary to the
case of anomalously close levels of opposite parity in rare-earth-metal atoms,
the natural linewidths of all the levels in the electronic ground state of a molecule are negligible
\cite{Brownbook2003}.
Most often, the ground state
of a molecule which has an odd number of electrons is either the  $^2\Sigma$ or the  $^2\Pi_{1/2}$ state.
However, since the rotational structure of the lower energy levels is simpler in the former case,
we will concentrate on  $^2\Sigma$ diatomics, which are the molecular equivalent of alkali-metal atoms.

Almost all the molecules that have $^2\Sigma$ as their ground state pertain to the Hund's $b$ coupling scheme. Following
Ref. \cite{Brownbook2003}, ${\bf{L}}$ (electronic orbital angular momentum) is coupled to ${\bf{R}}$
(rotational angular momentum of the nuclei) to form ${\bf{N}}$,
and ${\bf{N}}$ is coupled to ${\bf{S}}$ (electronic spin)
to form ${\bf{J}}$. Their corresponding projections on the internuclear axis, ${\bf{n}}$, are,
respectively, $m_{L}$, $m_{N}$, $m_{S}$ and $m_{J}$.
Including nuclear hyperfine interactions by the coupling ${\bf{J}}+{\bf{I}}={\bf{F}}$, one obtains
the spin-rotation-hyperfine Hamiltonian:
\begin{equation}
H_{\rm srh}=B\,{\bf{N}}^{2}+\gamma {\bf{N}} \cdot {\bf{S}}+b\, {\bf{I}}\cdot {\bf{S}} +c\,
({\bf{I}}\cdot {\bf{n}})({\bf{S}}\cdot {\bf{n}}),
\end{equation}
where $\gamma$ is the spin-rotation constant and $b$, $c$ are hyperfine constants.

In most cases of interest, $B \gg \gamma,b,c$, so $N$ is a good quantum number with eigenstates of energy
$E_{N}\simeq B\,N(N+1)$ and parity
$P=(-1)^{N}$. Since one can use a magnetic field to Zeeman shift consecutive states of opposite parity
($N=0,1$), and the magnetic field necessary
to overcome the rotational energy is large enough (about 1 T when the rotational constant is $B\sim 1$ GHz),
one can work in the decoupled basis
$|N,m_{N}\rangle|S,m_{S}\rangle|I,m_{I}\rangle$.

Within the subspace of rotational-hyperfine levels, the parity violating Hamiltonian (\ref{PV-Ham}) can
be written effectively \cite{Flambaum1985}
\begin{equation}
H_{\rm pv}=W_{p}\, \kappa\, ({\bf{n}} \times {\bf{S}}) \cdot {\bf{I}},
\end{equation}
where $W_{p}$ can be calculated using semiempirical methods \cite{Kozlov1985}.

The opposite-parity levels
\begin{eqnarray}
|A\rangle&\equiv& |0,0 \rangle|1/2,m_S \rangle |I,m_{I} \rangle \nonumber \\
|B\rangle&\equiv& |1,m_{N}' \rangle|1/2,-m_S\rangle |I,m_{I}' \rangle
\end{eqnarray}
can be mixed by $H_{\rm pv}$ when the quantum number $m_{F}=m_{N}+m_{S}+m_{I}$ is identical for
both states $|A\rangle$ and $|B\rangle$
\cite{Kozlov1995}.
Thus, once $W_{p}$ has been calculated \cite{Kozlov1985} together with the matrix elements
of the operator
$({\bf{n}} \times {\bf{S}})\cdot {\bf{I}}$ between opposite-parity states, we seek to determine $\kappa$ by measuring
\begin{equation}
iW\equiv \kappa W_{p} \langle A|({\bf{n}} \times {\bf{S}}) \cdot  {\bf{I}}| B\rangle.
\end{equation}
Importantly, time reversal invariance ensures that $iW$ is purely imaginary.
We note that this scheme is currently being designed to measure $iW$ using a Stark-interference technique \cite{DeMille2008}.

Here we focus on the effect of $H_{\rm pv}$ in an internal
bosonic Josephson junction (BJJ) between two rotational states in a gas of Bose condensed diatomic molecules,
achievable with foreseeable advances in molecular laser cooling.
In this case, PV effects can be manifested in the population difference and in the relative phase between the two modes, which can be
directly measured both in the Rabi and Josephson regimes \cite{Gross2010}.
The description of the Gross-Pitaevskii (GP) dynamics reduces, at low energies, to a nonlinear two-mode equation for the time-dependent
amplitudes, $\psi_{i}(t)=\sqrt{N_{i}(t)}\,e^{i\theta_{i}}$ ($i=A,B$), where $N_{i}$ and $\theta_{i}$ are,
respectively, the condensate molecule number and phase in state $i$. When the two condensates are coherently linked,
we write the wave function as
\begin{equation}
|\Psi(t)\rangle=\psi_{A}(t)|A\rangle +\psi_{B}(t)|B\rangle,
\end{equation}
where $N_{A}(t)+N_{B}(t)=N_{T}$.

Including $H_{\rm pv}$, the amplitudes and phases obey the GP equation 
\begin{equation}
i \frac{\partial}{\partial t}\left[
\begin{array}{c}
\psi_{A}\\
\psi_{B}
\end{array}
\right]=
\left[
\begin{array}{ccc}
E_{A} & -\Omega/2+iW \\
-\Omega/2-iW & E_{B} \\
\end{array}
\right]
\left[
\begin{array}{c}
\psi_{A} \\
\psi_{B}
\end{array}
\right]
\end{equation}
where $E_{i}$ is the effective energy (which includes the mean-field contribution), and $\Omega$ is the highly tunable Rabi
frequency connecting the two modes
\cite{Smerzi1997,Zapata1998,Cirac1998}.

Defining the fractional population imbalance, $z(t)\equiv[N_{A}(t)-N_{B}(t)]/N_{T}$, the relative phase,
$\phi(t)\equiv \theta_{A}(t)-\theta_{B}(t)$, and rescaling to a dimensionless time
$\Omega\, t\rightarrow t$,
we obtain
\begin{eqnarray}
\label{eqs}
\dot{z}&=&-\sqrt{1-z^{2}}\left( \sin \phi+w\cos \phi\right) \nonumber \\
\dot{\phi}&=&\Lambda \,z+\varepsilon \\
&+& \frac{z}{\sqrt{1-z^{2}}}
\left(\cos \phi-w\sin \phi\right) \nonumber,
\end{eqnarray}
where $\varepsilon$ is the effective detuning, $\Lambda \equiv U\,N_{T}/\Omega$, $w\equiv 2W/\Omega$ and
$U$ encompasses the various interactions.
We note that these equations represent a generalization of the well known pendulum
equations that describe the classical dynamics of two weakly coupled Bose-Einstein condensates
\cite{Smerzi1997,Zapata1998}.
They clearly show the PV effect in a molecular BJJ, with the parity violating contribution $W$ competing with
the usual coupling matrix element $\Omega/2$. We conclude that electroweak effects will have measurable consequences on the macroscopic
dynamics of the condensate.

Equations (\ref{eqs}) derive from the Hamiltonian
\begin{equation}
\label{eqham}
H=\frac{\Lambda z^{2}}{2}+\varepsilon z -\sqrt{1-z^{2}}\left(\cos \phi-w\sin \phi\right),
\end{equation}
which describes a {\it parity-violating non-rigid pendulum}, where the parity operation in
this representation changes $\phi \rightarrow -\phi$.
Within this mechanical analog, PV is reflected in the shifted equilibrium position of the
pendulum, which now is $\phi_w=\arctan w$ due to the $\sin \phi$ term in (\ref{eqham}).

For $\varepsilon=0$ and negligible interactions ($\Lambda=0$), the oscillations have the modified Rabi frequency
\begin{equation}
\bar \Omega = \Omega\sqrt{1+w^{2}}.
\end{equation}
In this simple case, the population imbalance evolves as
\begin{equation} \label{z-t}
z(t)=\sqrt{\frac{A_{1}}{1+w^{2}}}
\,\sin (\bar \Omega t+A_{2}) ,
\end{equation}
where $A_{1}=1+w^{2}-H^{2}$ and
$A_{2}=\sin^{-1}\left[z_{0}\sqrt{(1+w^{2})/A_{1}}\right]$, with $z_{0}\equiv z(0)$. The evolution of $\phi(t)$ is also very
sensitive to $iW$ but cannot be written in closed form.

\begin{figure}[ht!]
\includegraphics[width=.40\textwidth,height=.40\textwidth]{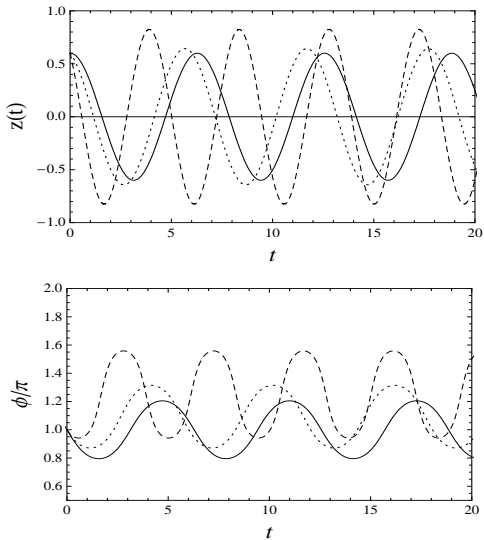}
\caption{\label{fig1}Time evolution of the population imbalance and phase difference for $\Lambda=0$ and
$w=0$ (solid), 0.3 (dotted) and 1.0 (dashed). We have taken $z_{0}=0.6$ and $\phi_{0}=\pi$.}
\end{figure}

Deviation from the standard non-rigid pendulum dynamics is shown in the time dynamics of both the population imbalance
and phase difference (Fig. \ref{fig1}).

In the presence of interactions \cite{comment2}, the harmonic limit ($|z|,|\phi-\phi_w| \ll 1$) displays sinusoidal oscillations of
$z(t)$ with a modified Josephson-Rabi frequency
\begin{equation}
\bar \omega_{\rm JR}=\bar \Omega\, (1+\bar\Lambda)^{1/2} \, ,
\end{equation}
where $\bar\Lambda=\Lambda/\sqrt{1+w^2}$ is the PV-renormalized interaction parameter.
The linearized Eqs. (\ref{eqs}) read
\begin{eqnarray}
&&\ddot z+ \bar \omega_{\rm JR}^{2} z=0 \nonumber \\
&&\ddot \phi+\bar \omega_{\rm JR}^{2}(\phi-\phi_w)=0.
\end{eqnarray}
We conclude that in the harmonic limit, PV effects change the Josephson-Rabi frequency.
In the non-interacting limit, the physics is formally identical to that of a non-degenerate 
gas of molecules, if the external fields are uniform. In the presence of inhomogeneities, 
an important advantage of the condensate is that it permits an identical response from all molecules, which should 
improve the accuracy of the experiment. Following Ref. \cite{Williams2000}, we estimate that the condensate 
response to a spatially varying Rabi frequency in the
$z$ axis is rigid when $E_{1}-E_{0}> |\Omega'| \langle 1|z|2 \rangle$, where $\Omega'$ stands for the gradient of the
Rabi frequency, and $1$ and $2$ refer to the two lowest states of the trap.
Taking typical values for trapping frequencies of $2\pi \times$5 Hz and trap sizes of about 25 $\mu$m, we obtain that 
the condensate response is rigid when $|\Omega'|< 2\pi \times 0.1$ Hz $\mu$m$^{-1}$. 

\begin{figure}[ht!]
\includegraphics[width=.40\textwidth,height=.25\textwidth]{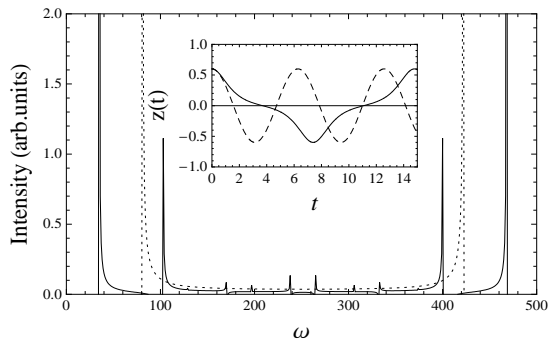}
\caption{\label{fig4}Time evolution of the population imbalance (inset) and its Fourier transform
for $\Lambda=\Lambda_{c}=1.11$ and $w=0$ (solid) and 0.2 (dotted).}
\end{figure}

In the strongly nonlinear regime, macroscopic quantum self trapping (MQST) is also present
\cite{Smerzi1997}. In fact,
the value $z_{0}=0$ is inaccesible at any time if
\begin{equation}
\Lambda > \bar \Lambda_{c}=\Lambda_{c}+\frac{w}{z_{0}^{2}}
\sqrt{1-z_{0}^{2}}\sin \phi_{0},
\end{equation}
where $\phi_0 \equiv \phi(0)$ and $\Lambda_{c}=\frac{2}{z_{0}^{2}}\left(\sqrt{1-z_{0}^{2}}\cos \phi_{0}+1 \right)$
is the critical parameter for MQST in absence of PV effects.
In Fig. \ref{fig4} we show, for $\Lambda=\Lambda_{c}=1.11$, the comparison of parity-even
and parity-odd effects in the Fourier
transform of the population imbalance, noting that the PV effects are easily
distinguishable from their parity conserving counterpart. We notice that the harmonic content of $z(t)$ is quite sensitive to $iW$ for large $\Lambda$. In the inset we
explicitly show the effect of PV in the time evolution of the population imbalance. We note that,
due to the difference between $\bar \Lambda_{c}$
and $\Lambda_{c}$, the onset of MQST depends on the value of $w$.

Once an appropriate control over the internal states of diatomic molecules is achieved, one also needs a precise measurement
 of the conjugate variables $z,\phi$  \cite{Gross2010} to measure PV effects in a molecular BJJ setup. Hereafter we focus on the molecule $^{87}$Rb\,$^{{\it A}}$Yb (with even $A$).
Ytterbium is so far one of the few elements besides
the alkali metals that can be trapped and condensed in optical traps \cite{Takasu2003}. In addition, as Yb is a
closed $s$-shell atom with zero nuclear spin, the only active electron in the molecule is the Rb $s$-electron.
Thus, from the point of view of the electroweak interaction, the Yb atom plays a passive role.

First we estimate the value of $\kappa$. $^{87}$Rb has $I=3/2$ with an
unpaired proton in the $p$ state. Thus, $\kappa=\kappa_{a}+\frac{5}{4}\kappa_{2}-\frac{15}{8}\kappa_{Q}$.
To get an insight on the order of magnitude expected for the parity violating effect, we consider $\kappa \simeq \kappa_{a}$
 \cite{Haxton2001,Sheng2010}. Taking $g_{p}\simeq 5$ \cite{Haxton2001},
we get 100$\kappa_{a}\simeq$ 27.
Given the proportionality between matrix elements of different weak interactions, we estimate the value of $W_{p}$ using previous 
molecular electronic structure calculations of the electron electric dipole moment of RbYb \cite{Meyer2009}, getting $W_{p}\simeq 36$ Hz 
(for comparison, we note that RaF has one of the largest values for
$W_{p}$ predicted so far, which is in the range of kHz \cite{Isaev2010}).
Thus, we conclude that the extra coupling that has to be accounted for in a RbYb-BJJ is, including only the anapolar
term, $W\simeq 4 \, \mathrm{Hz}$. If we include also the contribution of $\kappa_{2}\simeq0.05$ and $\kappa_{Q}\simeq0.01$ \cite{Sheng2010},
we obtain $W\simeq 5$ Hz. 

The Rabi frequency $\Omega$ is a highly tunable parameter because of its linear dependence on the applied 
dc electric field. We note that RbYb has a typical value for the electric dipole matrix element of 1 kHz cm V$^{-1}$. 
Thus, for instance, a value $\Omega \sim 100$ Hz can be obtained using an electric field $\sim $ 0.1 V cm$^{-1}$.
Due to the small value of this matrix element, 
dipole-dipole interactions are negligible compared to both $\Omega$ and W. In addition, we note that Zeeman 
degeneracy can be controlled up to $10^{-2}$ Hz for RbYb due to the high level of magnetic
field control (few tens of $\mu$G) recently achieved in a cold atom experiment \cite{Smith2011}. 

The curves $z(t)$ [Eq. (\ref{z-t})] and $\phi(t)$ (which lacks a closed form but can be calculated numerically with arbitrary accuracy) 
depend on $w=2W/\Omega$ through their frequency and amplitude. If we focus on the Rabi regime, where interactions are negligible, and assume 
that we estimate the value of $w$ only from the measurement of the oscillation frequency, the resolution obtained for $W\simeq 5$ Hz is 
known to be $\Delta W \sim f/(\tau\sqrt{N_{T}})$, 
where $\tau$ is the coherence time and $f$ is numerical factor taking into account recent experimental capabilities in
the absorption imaging of ultracold molecules. We take $N_{T}\simeq 4\cdot 10^{4}$ as the total number of condensed molecules and 
assume $\tau \sim$ 100 ms. Following Ref. \cite{Wang2010} we take $f=20$ and get that the  effect of $H_{\rm pv}$ could be experimentally 
detected within a relative precision of $\sim 20\%$.
We point out that this improvement in the accuracy of $\Delta W$ is due to the long coherence times of trapped cold molecules, 
as compared to those of a molecular beam experiment
\cite{Isaev2010}.
Concerning this coherence time, an important point is the role of losses. For RbYb we expect a loss time of about 
100 ms \cite{DOI}. 
This time should be larger than both the Rabi and
PV times, $\Omega^{-1}$ and $W^{-1}$, for
this proposal to be viable. 
The value of $\Omega$ is not a problem since it is highly tunable (or even zero). 
For RbYb, taking $W\simeq 5$ Hz, this condition can hardly be satisfied. However, we note that current estimates of $W_{p}$ are still rather
crude and the real situation might turn out to be more favourable. On the other hand, molecules such RaF, HgH or HgF are known to have
much larger values of $W_{p}$ \cite{Isaev2010}.

Understanding the final sensitivity that could be
reached by using a molecular BJJ requires a careful study of systematic effects.
As noted in Refs. \cite{Nguyen1997,DeMille2008}, care has to be taken of stray electric fields
that could mimic the PV effect. Due to the small value of the electric dipole matrix elements, it suffices to control
stray fields within a feasible accuracy of 0.1 V cm$^{-1}$.
Thermal effects can be neglected if the temperature is smaller than a typical rotational level splitting ($\sim$ mK). However, temperature can
be bigger than $W$ and $\Omega$, thanks to the fine control of the initial state.


We finally note that in the proposed setup a large number of curves depend on a few parameters. This intrinsic redundancy will permit not 
only a consistency check of the underlying physics but a precise measurement of the anapole moment
in a given molecule. The eventual measurement of this moment in a variety of molecules may ultimately reveal the strength of the poorly known
electron couplings to up and down quarks.

We acknowledge useful discussions with A. J. Leggett, I. Zapata and A. Dorta-Urra. This work has been supported by MICINN (Spain) 
through Grants No. FIS2007-65723, No, FIS2010-21372, and No. CTQ2008-02578/BQU, and the Juan de la Cierva program (P. B.).


\begin{thebibliography}{n}
\bibitem{comment2} We assume $\Lambda \ll N^{2}_{T}$ so that global phase coherence and thus the validity of the GP equation is guaranteed.
\end{thebibliography}
\end{document}